\documentclass[3p,times,twocolumn]{elsarticle}

\usepackage{ecrc}

\volume{00}

\firstpage{1}

\journalname{Journal A}

\runauth{Adriano}

\jid{procs}

\jnltitlelogo{Journal A}

\CopyrightLine{2011}{Published by Elsevier Ltd.}

\usepackage{amssymb}
\usepackage{amsmath}
\usepackage[figuresright]{rotating}

\usepackage{subfig}

\usepackage{graphicx}% Include figure files
\usepackage{dcolumn}% Align table columns on decimal point
\usepackage{bm}% bold math
\usepackage{epstopdf}
\usepackage{epsfig}
\usepackage[colorlinks = true,
linkcolor = blue,
urlcolor  = blue,
citecolor = blue,
anchorcolor = blue]{hyperref}% add hypertext capabilities
\usepackage{url}
\usepackage[mathlines]{lineno}% Enable numbering of text and display math
\usepackage{amssymb}
\usepackage{amsthm}
\usepackage{ulem}

\usepackage[figuresright]{rotating}
\usepackage{subcaption}

\usepackage{anyfontsize}

\usepackage{graphicx}% Include figure files
\usepackage{dcolumn}% Align table columns on decimal point
\usepackage{bm}% bold math
\usepackage{epstopdf}
\usepackage{epsfig}
\usepackage[colorlinks = true,
linkcolor = blue,
urlcolor  = blue,
citecolor = blue,
anchorcolor = blue]{hyperref}% add hypertext capabilities
\usepackage{url}
\usepackage[mathlines]{lineno}% Enable numbering of text and display math
\theoremstyle{remark}

\usepackage{xcolor}

\newcommand{\dd}{\mathrm{d}}

\begin{document}

% \nolinenumbers

\begin{frontmatter}

%% Title, authors and addresses

%% use the tnoteref command within \title for footnotes;
%% use the tnotetext command for the associated footnote;
%% use the fnref command within \author or \address for footnotes;
%% use the fntext command for the associated footnote;
%% use the corref command within \author for corresponding author footnotes;
%% use the cortext command for the associated footnote;
%% use the ead command for the email address,
%% and the form \ead[url] for the home page:
%%
%% \title{Title\tnoteref{label1}}
%% \tnotetext[label1]{}
%% \author{Name\corref{cor1}\fnref{label2}}
%% \ead{email address}
%% \ead[url]{home page}
%% \fntext[label2]{}
%% \cortext[cor1]{}
%% \address{Address\fnref{label3}}
%% \fntext[label3]{}

\dochead{}
%% Use \dochead if there is an article header, e.g. \dochead{Short communication}
%% \dochead can also be used to include a conference title, if directed by the editors
%% e.g. \dochead{17th International Conference on Dynamical Processes in Excited States of Solids}

\title{Is there an accelerating nonspreading wave packet in the Schr\"odinger equation with higher even-order dispersions?}

%% use optional labels to link authors explicitly to addresses:
%% \author[label1,label2]{<author name>}
%% \address[label1]{<address>}
%% \address[label2]{<address>}

		\author{Farrell Theodore Adriano\corref{cor1}}
		\ead{ftheodoreadriano@gmail.com}
        \author{Hadi Susanto}
		\ead{hadi.susanto@yandex.com}

\address{Department of Mathematics, Khalifa University, PO Box 127788, Abu Dhabi, United Arab Emirates}
\cortext[cor1]{Corresponding author}

\begin{abstract}
%% Text of abstract
Probably not. For Airy-type initial data, although nearly all lobes accelerate, the principal lobe remains essentially stationary. Moreover, the overall wave packet gradually disperses.
\end{abstract}

\begin{keyword}
%% keywords here, in the form: keyword \sep keyword

%% PACS codes here, in the form: \PACS code \sep code

%% MSC codes here, in the form: \MSC code \sep code
%% or \MSC[2008] code \sep code (2000 is the default)

\end{keyword}

\end{frontmatter}

%%
%% Start line numbering here if you want
%%
% \linenumbers

%% main text

\section{Introduction} \label{sec:intro}

Nonspreading wave packets are solutions of dispersive wave equations whose
probability density preserves its spatial profile in time, up to translation and phase modulation. For the free Schr\"{o}dinger equation,
such behavior is highly nontrivial because generic localized wave packets
broaden under the usual quadratic dispersion. In a seminal work, Berry and Balazs
\cite{Berry1979} showed that the one-dimensional free Schr\"{o}dinger
equation nevertheless admits a remarkable exact solution: the Airy wave
packet. Its intensity profile translates along a parabolic trajectory,
exhibiting self-acceleration in the absence of any external force.

This behavior can be interpreted semiclassically as the
coherent superposition of classical trajectories whose interference
suppresses dispersion. Berry and Balazs further demonstrated that, up to
trivial symmetries, this accelerating nonspreading solution is unique for
quadratic dispersion and that the nonspreading property persists only in
the presence of at most a time-dependent linear potential
\cite{Berry1979,Lin2008}. Finite-energy realizations of Airy beams were
subsequently demonstrated experimentally in one and two transverse
dimensions \cite{Siviloglou2007}. The trajectory of such beams can be
controlled through phase engineering at the input plane
\cite{siviloglou2008ballistic,hu2010optimal}, and their self-reconstruction
property has been verified experimentally \cite{broky2008self}.
Extensions to higher-dimensional configurations have also been realized
\cite{chong2010airy,abdollahpour2010spatiotemporal}. For reviews, see,
e.g., \cite{mazilu2010light,zhang2017guided,efremidis2019airy,ren2021non}.

The existence of the Airy packet is closely tied to the quadratic form of
the dispersion relation. This raises a natural structural question: how do
wave-packet dynamics depend on the order of the dispersive operator?
Higher even-order dispersion is not merely of formal interest. In
dispersion-engineered optical media, the usual second-order
group-velocity dispersion can be suppressed, allowing fourth-order or
higher even-order terms to dominate the propagation dynamics.

Recent experiments have shown that when fourth-order dispersion becomes
dominant, it can balance Kerr nonlinearity to produce localized pulses
whose structure and scaling properties differ substantially from those of
conventional quadratic-dispersion solitons
\cite{blanco2016pure,runge2020pure,wang2026revealing,
runge2021infinite,han2024pure}. More generally, nonlinear models with
dominant higher even-order dispersion support a wide variety of localized
states with distinct tail behavior, dynamical symmetries, and scaling laws
\cite{tam2019stationary,wu2025pulsating,silvestri2025pure,liu2021raman,
wang2022raman,zhu2025raman,alexander2022dark,li2025interactions,
parra2024pure,li2025pure,deng2025multiple,deng2025internal,
bandara2021infinitely,tsolias2023kink}. In particular, such systems
generally lack Galilean invariance \cite{widjaja2021absence}, so moving localized states do not arise through simple boost transformations.

While these developments highlight the richness of nonlinear phenomena associated with higher-order dispersion, a fundamental linear question remains open. Does a purely dispersive Schr\"{o}dinger equation with dominant quartic (or higher even-order) dispersion admit a nonspreading wave packet that accelerates in free space, analogous to the Berry--Balazs Airy solution in the quadratic case? Is self-accelerating shape-preserving propagation a generic feature of even-order dispersion, or is it specific to the quadratic operator?

In this work, we show that it is the latter. For the Schr\"{o}dinger equation with pure quartic dispersion, we show that no globally shape-preserving solutions undergoing accelerated motion exist.
For Airy-type initial data, the principal lobe remains near its initial position, while the whole wave packet slowly disperses. The self-accelerating Airy phenomenon is therefore specific to quadratic dispersion and does not extend to the pure-quartic case. We further extend this conclusion to
higher even-order dispersive operators.

\section{Analysis of Quartic Dispersion}
\label{sec:analysis}

\subsection{Nonexistence of Rigidly Accelerating Shape-Preserving Solutions}

We consider the fourth-order Schr\"odinger equation
\begin{equation}
    i \partial_t u = (\partial_x^4 + V(x,t)) u.
    \label{eq:4th_order_schrodinger}
\end{equation}
Our goal is to determine whether this equation admits
shape-preserving solutions undergoing accelerated motion.
We begin with the natural analogue of the Airy construction.

Let $\mathrm{Ai}_4(x)$ denote a hyper-Airy solution of the fourth-order ordinary differential equation
\begin{equation}
    \mathrm{Ai}_4^{(4)} - x \mathrm{Ai}_4 = 0,
    \label{eq:hyperairy_eqn}
\end{equation}
which plays the role of the Airy function for quartic dispersion.
We consider the ansatz
\begin{equation}
    u(x,t) = \mathrm{Ai}_4(x-d(t)) e^{i\phi(x,t)},
    \label{eq:ansatz_nonspreading_hyperairy}
\end{equation}
where $d(0)=0$ and $\phi(x,0)=0$.
If such a solution were nonspreading, then $|u(x,t)|$
would coincide with $|\mathrm{Ai}_4(x)|$ up to translation.

Differentiating \eqref{eq:ansatz_nonspreading_hyperairy} and substituting
into \eqref{eq:4th_order_schrodinger}, we obtain
\[
i\partial_t u - \partial_x^4 u
=
\sum_{j=0}^{3}
A_j(x,t)\,
\mathrm{Ai}_4^{(j)}(x-d(t))\,e^{i\phi(x,t)},
\]
where the coefficients $A_j$ depend on $\phi$, $d$, and $V$.
The explicit computation yields
\begin{subequations}
\begin{align}
A_3 &= -4i\phi_x, \label{eq:A3}\\
A_2 &= -6i\phi_{xx} + 6\phi_x^2, \label{eq:A2}\\
A_1 &= -i\dot d -4i\phi_{xxx}
      +12\phi_x\phi_{xx}
      +4i\phi_x^3, \label{eq:A1}\\
A_0 &= -\phi_t
      -i\phi^{(4)}
      +4\phi_x\phi^{(3)}
      +3\phi_{xx}^2
      +6i\phi_x^2\phi_{xx}\nonumber\\
      &-\phi_x^4
      -x + d - V. \label{eq:A0}
\end{align}
\end{subequations}

Since $\{\mathrm{Ai}_4,\mathrm{Ai}_4',\mathrm{Ai}_4'',\mathrm{Ai}_4'''\}$
form a fundamental system of solutions of
\eqref{eq:hyperairy_eqn}, they are linearly independent.
Therefore, each coefficient $A_j$ must vanish identically.

The vanishing of these coefficients leads to the following conclusions. From \eqref{eq:A3}, we obtain $\phi_x = 0$, so $\phi$ depends only on time $\phi(x,t) = \phi(t)$. Substituting this into \eqref{eq:A2} shows that $A_2$ vanishes automatically. Consequently, the vanishing of $A_{1}$ implies that $\dot{d} = 0$, i.e. the translation $d(t) = c$ is constant in time. In particular, accelerated motion is impossible for the $\mathrm{Ai}_{4}$ wavepacket. Finally, \eqref{eq:A0} simplifies to
\[
-\phi_t - x + c - V(x,t)=0,
\]
that can hold only if
\[
V(x,t) = -x
\quad \text{and} \quad
\phi(t)=ct.
\]
Thus, $\mathrm{Ai}_4$ is stationary in the linear potential $V=-x$, but no accelerated translation occurs. We, therefore, conclude that Eq.~\eqref{eq:4th_order_schrodinger} does not admit rigidly accelerating shape-preserving solutions of the form \eqref{eq:ansatz_nonspreading_hyperairy}.

While the above analysis shows that the $\mathrm{Ai}_{4}$ wavepacket does not accelerate while preserving its shape under \eqref{eq:4th_order_schrodinger}, we conjecture that there are no such wavepackets in the fourth order Schr\"{o}dinger equation. This may be seen through the following heuristic observations.

To exclude the possibility that some other profile might accelerate, we consider the more general ansatz
\begin{equation}
    u(x,t)=f(x-d(t)) e^{i\phi(x,t)},
    \label{eq:general_ansatz}
\end{equation}
where the shape profile $f$ satisfies
\begin{equation}
    (\partial_x^4+V_0(x))f=E_f f,
\end{equation}
for some potential $V_{0}(x)$ and scalar $E_{f}$.
Substitution into \eqref{eq:4th_order_schrodinger} and collecting the term $(i\partial_{t} - \partial_{x}^{4})u$ to one side, we obtain
\begin{equation}
    \begin{split}
        (i\partial_{t}-\partial_{x}^{4})u = & [-4i\phi_{x}\partial_{x}^{3} - (6i\phi_{xx}+6\phi_{x}^{2})\partial_{x}^{2} \\
        -&(i\dot{d}+4i\phi_{xxx}+12\phi_{x}\phi_{xx}-4i\phi_{x}^{3})\partial_{x}\\
    -&(E_{f} - V_{0}(q) + i\phi_{xxxx} + 4\phi_{x}\phi_{xxx} \\
    +&3\phi_{xx}^{2}-6i\phi_{x}^{2}\phi_{xx}-\phi_{x}^{4} + \dot{d}\phi_{x}+\phi_{t})]u,
    \end{split}
\end{equation}
with $q = x-d(t)$. We now identify the operator acting on $u$ as the potential $V(x,t)$.
If we restrict to physical potentials $V(x,t)$
that only act as multiplication operators,
these derivative terms must vanish,
which forces $\phi_x=0$.
Consequently $\dot d(t)=0$ as before.
Hence, no rigidly accelerating shape-preserving solutions exist.

The nonexistence of an accelerating nonspreading wave packet in the free fourth-order Schr\"{o}dinger equation, \eqref{eq:4th_order_schrodinger} with $V(x,t) = 0$ may also be seen by considering a decomposition of the corresponding time evolution operator $e^{-it p^{4}}$, with $p = -i \partial_{x}$ denoting the momentum operator. Following \cite{Unnikrishnan1996}, a nontrivial decomposition of $e^{-itp^{4}}$ of the form
\begin{equation} \label{eqn:operator_decomp}
    e^{-it p^{4}} = e^{i\delta(t)}e^{i\gamma(t)g(x)}e^{id(t)p}e^{iv(t)H(x,p)}
\end{equation}
for some $\delta(t)$, $\gamma(t)$, $g(x)$, $f(t)$, and $H(x,p)$ (such that $\delta(0) = \gamma(0) = f(0) = 0$) would allow for an accelerating shape-preserving wavepacket since an eigenfunction $u(x,0) = f(x)$ of $H$ satisfying $H f = E f$ would evolve as
\begin{align*}
    u(x,t) = e^{i\delta(t)}e^{i\gamma(t)g(x)}e^{i Ev(t)}f(x-d(t)).
\end{align*}
However, a decomposition of the form \eqref{eqn:operator_decomp} is not possible. If it were, then it must hold that
\begin{align} \label{eqn:operator_decomp_2}
    e^{iv(t)H(x,p)} = e^{-i\delta(t)}e^{-id(t)p}e^{-i\gamma(t)g(x)}e^{-itp^{4}}.
\end{align}
Taking the logarithmic derivative of \eqref{eqn:operator_decomp_2} with respect to time gives that
\begin{equation} \label{eqn:H}
    H = -\frac{1}{\dot{v}(t)}\left[\dot{\delta} + \dot{\eta}p + \dot{\gamma}g(x-d) + (p + \gamma g'(x-d))^{4}\right].
\end{equation}
For $H = H(x,p)$ to be time-independent, from the coefficient of the $p^{4}$ term in \eqref{eqn:H}, it follows that $\dot{v} = 0$, so that $v(t)$ is constant. Consequently, for the coefficient of $p^{3}$ in \eqref{eqn:H} to vanish, $\gamma(t)g'(x-d(t)) = 0$, which implies either $\gamma(t) = 0$ or $g' \equiv 0$, both of which lead to a trivial decomposition of $e^{-itp^{4}}$. Thus, if one were to pursue this direction to obtain the accelerating shape-preserving solution, one might need to consider different decompositions of the evolution operator $e^{-itp^{4}}$.

\subsection{Time Evolution of Hyper-Airy Initial Data}

We now analyze the free equation ($V=0$) with
\begin{equation}
    u(x,0)=\mathrm{Ai}_4(x).
    \label{eq:hyperairy_initial_condition}
\end{equation}

Taking the Fourier transform of \eqref{eq:hyperairy_eqn} yields
\[
\widehat{\mathrm{Ai}}_4(k)=e^{-ik^5/5}.
\]
Solving in Fourier space yields
\begin{equation}
    u(x,t)
    =
    \frac{1}{2\pi}
    \int_{-\infty}^{\infty}
    \exp\!\left[
        i\Phi(k;x,t)
    \right] dk,
    \label{eq:integral_representation_solution}
\end{equation}
where the phase is 
\begin{equation}
\Phi(k;x,t)
=
-\frac{k^5}{5}
- t k^4
+ k x .
\label{eq:phase}
\end{equation}

\begin{figure}[tbhp]
    \centering
    \includegraphics[width=0.5\textwidth]{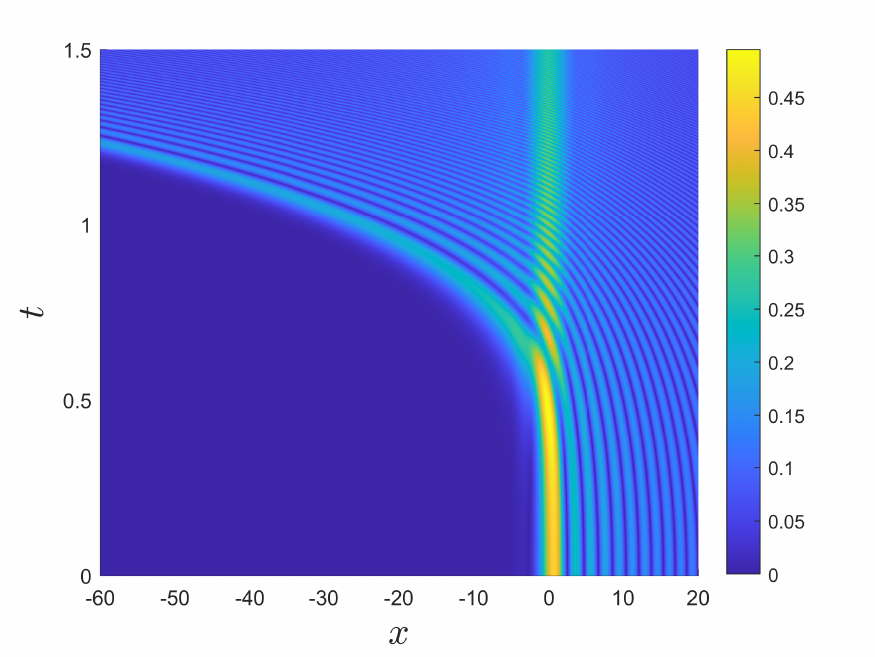}
    \caption{Propagation of $\mathrm{Ai}_4(x)$ under the free fourth-order Schr\"odinger equation, i.e., Eq.~\eqref{eq:4th_order_schrodinger} with $V(x,t)=0$. An accelerating caustic forms along $x=-27t^4$, while the amplitude decays dispersively.}
    \label{fig:time_dynamics_abs(psi)}
\end{figure}

Setting
\[
z = - \frac{k - t}{5^{1/5}},
\]
the oscillatory integral representation \eqref{eq:integral_representation_solution} can be reduced to the canonical swallowtail form \cite{NIST:DLMF}
\begin{equation}
u(x,t)
=
(-5)^{1/5}
e^{\,i\left(-t x - \frac{4}{5} t^{5}\right)}
\frac{1}{2\pi}
\mathrm{Sw}(a,b,c),
\label{eqn:solution_closed_form}
\end{equation}
where
\begin{equation}
\mathrm{Sw}(a,b,c)
=
\int_{-\infty}^{\infty}
e^{\,i\left(z^{5} + a z^{3} + b z^{2} + c z\right)} dz,
\end{equation}
with parameters
\begin{equation*}
a = - 2 \cdot 5^{3/5} t^{2},
\quad
b = - 4 \cdot 5^{1/5} t^{3},
\quad
c = - 5^{1/5} (3 t^{4} + x).
\end{equation*}
This representation is convenient both for asymptotic analysis
and for numerical evaluation.
Using \eqref{eqn:solution_closed_form}, the time evolution can be computed directly. As the integrand is highly oscillatory, the numerical computation of the integral is done in Matlab using the Pathfinder toolbox \cite{Gibbs2025}. This package implements the method of numerical steepest descent \cite{doi:10.1137/050636814} using the algorithm described in \cite{GIBBS2024112787}. 
% \textbf{\Large Please add more details how the integration was done. Maybe the name of the package?}
The resulting dynamics are shown in Fig.~\ref{fig:time_dynamics_abs(psi)}.

\subsection{Asymptotics Analysis}

From Fig.~\ref{fig:time_dynamics_abs(psi)}, it appears that the hyper-Airy wave packet accelerates along a curve. We will show below, that it is given by 
\begin{equation}
 x=-27t^{4}, \label{eq:caustic}   
\end{equation}
which defines a caustic. We will also show that its amplitude decreases with increasing $t$. 

In dispersive wave propagation, a caustic is the locus in physical space where wave energy concentrates due to the coalescence of propagation paths. Geometrically, it is the envelope of the associated Hamiltonian rays and marks the boundary between oscillatory regions and regions of rapid decay. In physical terms, caustics typically appear as intensity ridges or focusing curves \cite{berry1980iv,nye1999natural}.

In integral representations of wave fields, caustics arise when stationary points of the phase merge. At such points, the standard quadratic approximation of the phase fails, and higher-order terms determine the local structure of the solution. The resulting behavior exhibits characteristic amplitude scaling and profile transitions. 

For the present quartic-dispersion problem, consider the phase \eqref{eq:phase}. A caustic occurs when stationary points coalesce, i.e.,
when both
\[
\Phi'(k)=0
\quad \text{and} \quad
\Phi''(k)=0
\]
are satisfied, from which we obtain the nontrivial solution $k_* = -3t$. Substituting this into $\Phi'(k)=0$ yields the accelerating curve \eqref{eq:caustic}.

To derive the asymptotic behavior near this curve,
set
\[
k = k_* + \kappa.
\]
Expanding the phase around $k_*$ gives
\[
\Phi(k)
=
\Phi(k_*)
+ \frac{\kappa^3}{3!}\Phi^{(3)}(k_*) 
+ \frac{\kappa^4}{4!}\Phi^{(4)}(k_*) 
+ \kappa (x+27 t^4)
+ \cdots .
\]
Since $\Phi'(k_*)=\Phi''(k_*)=0$, the leading nonvanishing derivative is $\Phi^{(3)}(k_*)$, so the cubic term dominates in a neighborhood of the caustic.

A straightforward computation yields
\[
\Phi^{(3)}(k_*) = -36 t^{2},
\]
so after rescaling
\[
\kappa = (18)^{-1/3}t^{-2/3} z,
\]
the integral reduces to the canonical Airy form.
This leads to the leading order asymptotic approximation
\begin{equation}
u(x,t)
\sim
t^{-2/3}
e^{-i\left(\frac{162}{5}t^5+3t x\right)}
\mathrm{Ai}\!\left(
-\frac{x+27 t^4}{(18)^{1/3} t^{2/3}}
\right),
\label{eq:large_t_asymptotic_ai4_near_caustic_refined}
\end{equation}
valid for
\[
|x + 27t^4| \sim t^{2/3}.
\]
% \[
% |x+27 t^4| \ll t^{3/2}.
% \]
%\textbf{\Large Theo, please check: is it $t^{3/2}$ or $t^{2/3}$? I think it is valid for $|x + 27t^4| \ll t^{3/2}$, Pak. Of course it is also valid whenever $|x + 27t^4| \sim t^{2/3}$, which is a smaller region than $t^{3/2}$.}
In fact, the approximation \eqref{eq:large_t_asymptotic_ai4_near_caustic_refined} is valid on a larger region if we include the higher-order terms. This can be seen by defining a coordinate along the caustic as $\delta = x + 27t^4$ (such that $\delta = 0$ gives the caustic curve) and letting $z = \delta t^{-2/3}(18)^{-1/3}$. Then, we can write $u(x,t)$ in power series form as
\begin{eqnarray}
     & u(x,t) \sim t^{-2/3}e^{-i\left(\frac{162}{5}t^5+3t x\right)}\times\nonumber\\ 
    &\left[\mathrm{Ai}(-z) + \sum_{j=1}^{\infty} c_{j} t^{-5j/3}\mathrm{Ai}^{(4j)}(-z)\right].
\end{eqnarray}
From the asymptotics of $\mathrm{Ai}^{(j)}(z)$ for large $|z|$, we see that the power series is asymptotic if $|\delta| \ll t^{3/2}$, so that \eqref{eq:large_t_asymptotic_ai4_near_caustic_refined} is valid in the region 
\[
|x+27t^{4}| \ll t^{3/2}.
\]

The asymptotic formula
\eqref{eq:large_t_asymptotic_ai4_near_caustic_refined}
shows that the amplitude along the caustic decays at $O(t^{-2/3})$
as $t \to \infty$.
Moreover, to the right of the caustic, the solution is oscillatory,
since the argument of the Airy function in \eqref{eq:large_t_asymptotic_ai4_near_caustic_refined}
is negative, whereas to the left of the caustic, it decays exponentially.

\begin{figure}[tbhp]
\centering
\includegraphics[width=0.45\textwidth]{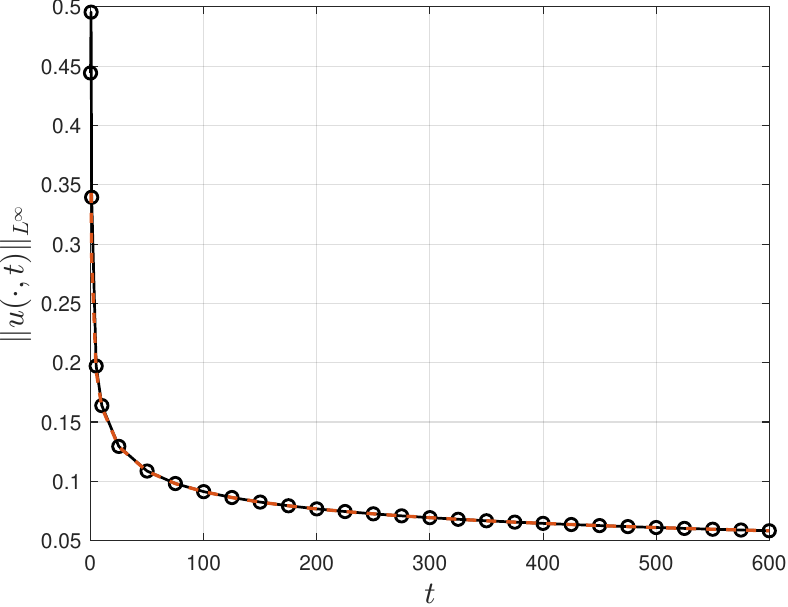}
\caption{
Time decay of $\|u(\cdot,t)\|_{L^{\infty}(\mathbb{R})}$.
Black circles: numerical evaluation.
Red dashed line: leading-order asymptotic prediction
from \eqref{eq:asymptotic_small_x}.
}
\label{fig:sup_against_t_ai4}
\end{figure}

Away from the caustic, further asymptotic regimes can be obtained
by stationary phase analysis. Analysis in those regimes can describe the dynamics of the principal lobe observed in Fig.~\ref{fig:time_dynamics_abs(psi)}. 

For small $x$ $(|x| \ll t)$, a uniform approximation of \eqref{eq:integral_representation_solution} may be obtained in terms of a Pearcey-type integral \cite{NIST:DLMF}. This is done by rescaling the variable $k = t^{-1/4}\xi$, letting $\zeta = xt^{-1/4}$, and rewriting the integral in \eqref{eq:integral_representation_solution}, denoted as $I(x,t)$ in a power series in $t^{-5/4}$ as 
\begin{equation}
    I(x,t) = t^{-1/4}\left[F(\zeta) + \sum_{j=1}^{\infty} \dfrac{(-1)^{j}}{5^{j}j!} t^{-5j/4}F^{(5j)}(\zeta)\right],
\end{equation}
where $F(\zeta) = \int_{-\infty}^{\infty} e^{-i(\xi^{4}-\zeta\xi)}\ \dd{\xi}$ is a Pearcey-type integral. Using the asymptotics of Pearcey-type integrals \cite{NIST:DLMF} for large $|\zeta|$, it is found that the power series is asymptotic as long as $|x| \ll t$. Therefore, in this region, the leading order behavior of $u(x,t)$ is given by the first term
\begin{equation}
    u(x,t) \sim t^{-1/4} \frac{1}{2\pi} F(\zeta), \quad |x| \ll t.
\end{equation}
Now, using the asymptotics of $F(\zeta)$ for small $|\zeta|$, in the central region $|x| \ll t^{1/4}$, one finds
\begin{eqnarray}
    \label{eq:asymptotic_small_x}
%    \begin{split}
    & u(x,t)
    \sim
    t^{-1/4}\frac{1}{2\pi}\times\nonumber\\
    & \Bigg[
        \frac{1}{2}e^{-i\pi/8}\Gamma\!\left(\frac{1}{4}\right)
        - \frac{1}{4}t^{-1/2}e^{-i3\pi/8}
        \Gamma\!\left(\frac{3}{4}\right)x^{2}
    \Bigg].
%    \end{split}
\end{eqnarray}
Thus, near the origin, the amplitude decays like $t^{-1/4}$. The $x^2$ term in \eqref{eq:asymptotic_small_x} explains the decaying lobe-like structure near $x \approx 0$ in Fig.~\ref{fig:time_dynamics_abs(psi)}.
For large $|\zeta|$ such that $t^{1/4} \ll |x| \ll t$, the asymptotics of $F(\zeta)$ yields the approximation
\begin{equation} 
    u(x,t)
    \sim
    \frac{1}{2^{5/6}\sqrt{3\pi}}
    t^{-1/6}|x|^{-1/3}
    e^{i\left(\frac{3}{4^{4/3}}t^{-1/3}|x|^{4/3}-\frac{\pi}{4}\right)}. \label{eq:asymptotic_intermediate_x}
\end{equation}

For other regions, we resort to stationary phase analysis of the integral. This is done by expanding the phase around its stationary points, $k_{*}$, as
\[
\Phi(k) = \Phi(k_{*}) + \frac{(k-k_{*})^{2}}{2!}\Phi''(k_{j}) + \dots.
\]
Then, the integral can be evaluated to give the leading order contribution of $k_{*}$ as
\begin{equation}
    I_{*}(x,t) = \sqrt{2\pi}|\Phi''(k_{*})|^{-1/2}e^{-i\pi/4}e^{i\Phi(k_{*})}.
    \label{eqn:leading_integral_stationary_phase}
\end{equation}
Stationary points of the phase satisfy
\begin{equation}  \label{eqn:stationary_phase_eqn}
\Phi'(k) = -(k^{4} + 4tk^{3} - x) = 0,
\end{equation}
which is a quartic in $k$. This quartic has discriminant $\Delta = -256x^2(x+27t^4)$. Thus, for $x > -27t^4$, there exist 2 real roots and 2 complex roots, no real roots when $x < -27 t^4$, and a double real root when $x = -27t^4$. Whenever real roots exist, they give dominant contributions to the integral in \eqref{eq:integral_representation_solution}, as the complex roots would account for exponentially small corrections to the integral.

In the region $t \ll |x| \ll t^{4}$, the two real stationary points are given asymptotically by
\begin{equation}
    k_{0} \sim \left(\frac{x}{4t}\right)^{1/3}, 
    % + O\left(t^{-1}\right), 
    \quad k_{1} \sim -4t.
    % + O\left(t^{-3}\right).
\end{equation}
Evaluating this at each stationary point gives that $I_{0}(x,t) = O(t^{-1/6}|x|^{-1/3})$ and $I_{1}(x,t) = O(t^{-3/2})$, thus $I_{0}$ is the dominant term, as $t \ll |x| \ll t^{4}$. This yields the same asymptotic approximation as \eqref{eq:asymptotic_intermediate_x}. Therefore, in the intermediate regime $t^{1/4} \ll |x| \ll t^{4}$, the asymptotics of $u(x,t)$ is given by \eqref{eq:asymptotic_intermediate_x}.
% \begin{equation}
%     \label{eq:asymptotic_intermediate_x}
%     u(x,t)
%     \sim
%     \frac{1}{2^{5/6}\sqrt{3\pi}}
%     t^{-1/6}|x|^{-1/3}
%     e^{i\left(\frac{3}{4^{4/3}}t^{-1/3}|x|^{4/3}-\frac{\pi}{4}\right)}.
% \end{equation}
Since in this regime $|x| \gtrsim t^{1/4}$,
we have $|x|^{-1/3} \lesssim t^{-1/12}$, and hence
\[
|u(x,t)| \lesssim t^{-1/4}.
\]

For larger $|x|$, the decay becomes stronger.
When $|x| \sim t^{4}$, dominant balance gives that the two real roots of \eqref{eqn:stationary_phase_eqn} scale as $k_{*}\sim t$. In view of \eqref{eqn:leading_integral_stationary_phase}, both are of equal contribution to the integral, as $|\Phi''(k_{*})|^{-1/2} \sim t^{-3/2}$ for both roots. Therefore, one obtains that $u(x,t)=O(t^{-3/2})$.

When $x \ll -27t^{4}$, there are no real stationary points.
In this region the leading contribution arises from the complex saddle points (stationary points). The leading order contribution can be computed via the steepest descent method in a similar way to the stationary phase analysis. The phase is expanded near the saddle points and the integral is evaluated. However, the integral contour is deformed to follow the phase's steepest descent contours. Doing so, it is found that asymptotically
\[
u(x,t)
=
O\!\left(
|x|^{-3/8}
e^{-2\sqrt{2}|x|^{5/4}/5}
\right),
\]
which implies exponential decay in $t$ since $|x|\ll t^{4}$.
When $x\gg t^{4}$, 
% \textbf{\Large Is it $\gg$ or $\ll$?}
the decay is algebraic since the two real roots of \eqref{eqn:stationary_phase_eqn} scale as $k_{*} \sim x^{1/4}$ by dominant balance.
The stationary-phase analysis then gives, at leading order, $u(x,t)=O(x^{-3/8})$.
Therefore, $|u(x,t)| \lesssim t^{-3/2}$ since $|x| \gg t^{4}$.

Combining these regimes, we conclude that
\[
\|u(\cdot,t)\|_{L^\infty(\mathbb{R})}
\sim t^{-1/4}
\quad \text{as } t \to \infty,
\]
with the slowest decay occurring in the region $|x| \ll t^{1/4}$.
This is consistent with the dispersive estimate for the free
fourth-order Schr\"odinger equation
\cite{Benartzi2000},
\begin{equation}
    \|u(\cdot,t)\|_{L^\infty(\mathbb{R})}
    \le
    C t^{-1/4}
    \|u(\cdot,0)\|_{L^1(\mathbb{R})}.
\end{equation}
even though the initial data $\mathrm{Ai}_4$ is not integrable.

The asymptotic prediction is compared with numerical evaluation of
$\|u(\cdot,t)\|_{L^\infty(\mathbb{R})}$ in
Fig.~\ref{fig:sup_against_t_ai4}.
The numerical solution is computed from
\eqref{eq:integral_representation_solution}
on the spatial interval $-60\le x\le 60$
with $t\in[0,600]$.
The agreement with the small-$x$ asymptotics
\eqref{eq:asymptotic_small_x}
improves as $t$ increases.

\begin{figure}[tbhp]
\centering
\includegraphics[width=0.45\textwidth]{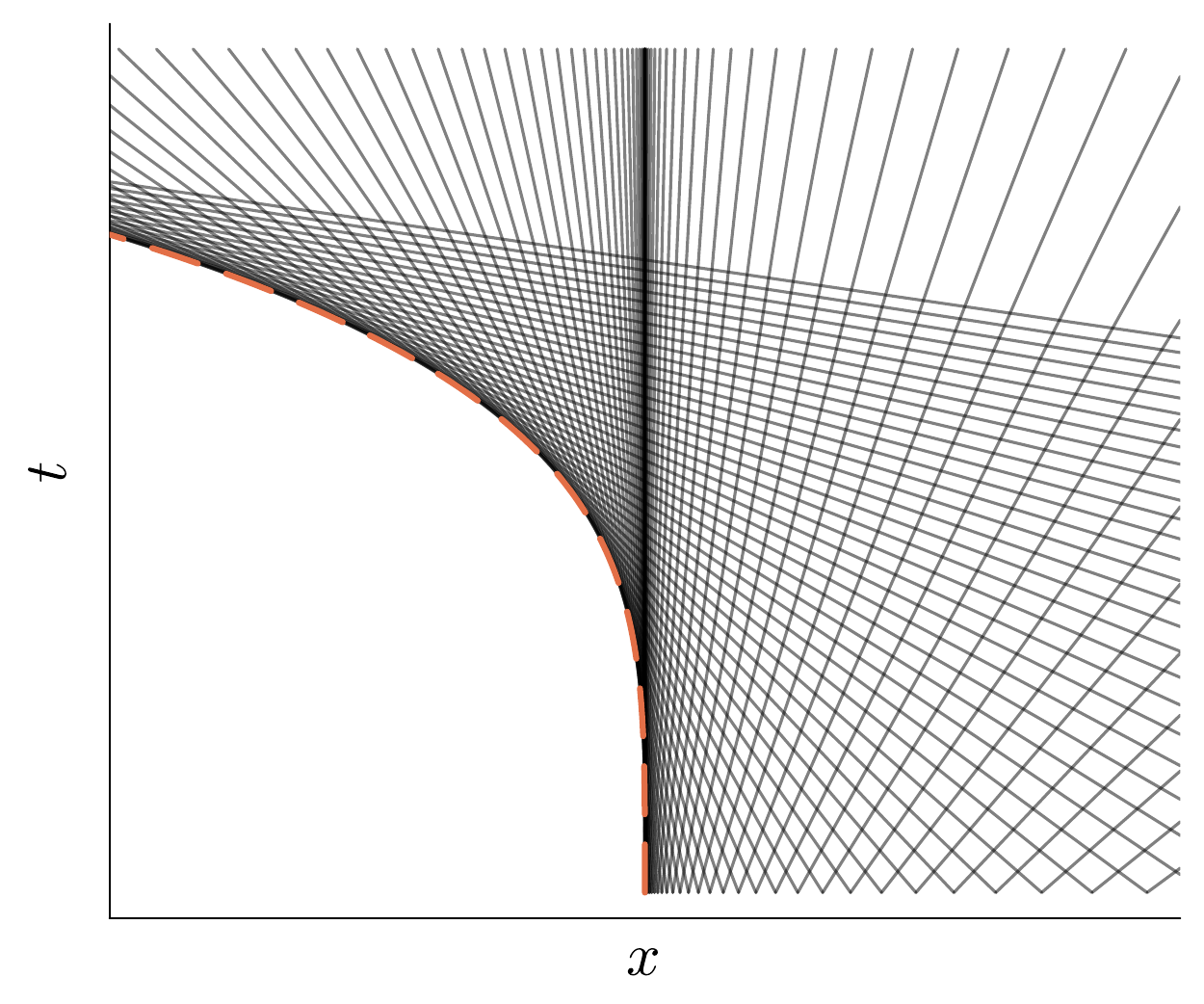}
\caption{Classical paths of the $\mathrm{Ai}_{4}$ packet in space time. The black lines are traced by the trajectories \eqref{eqn:classical_position}. The dashed red line is the quartic envelope given by the caustic \eqref{eq:caustic}.
}
\label{fig:classical_paths}
\end{figure}

%% Reviewer 4's comments about classical paths
It is also interesting to compare the time evolution of the $\mathrm{Ai}_{4}$ wavepacket in \eqref{eq:4th_order_schrodinger} with its classical counterpart, i.e., the classical paths traced by the packet under the Hamiltonian $H = p^{4}$ (here $p$ denotes the momentum operator). In the classical case, the packet is a family of orbits represented by a curve in the classical phase space as $p = P_{t}(x)$ at time $t$. The packet then evolves as each point in the curve obeys Hamilton's equations. For the $\mathrm{Ai}_{4}$ packet, the initial curve can be obtained via its asymptotic form as $x \gg 1$ as follows. It is found that as $x \gg 1$,
\begin{align}
    u(x,0) = \mathrm{Ai}_{4}(x) \sim \frac{1}{\sqrt{2\pi}}x^{-3/8}\sin\left(\frac{4}{5}x^{5/4} + \frac{\pi}{4}\right),
\end{align}
which follows the standard semiclassical form $u(x,0) \sim \left[(\partial_{x}^{2}S_{+})^{1/2}e^{iS_{+}x} + (\partial_{x}^{2}S_{-})e^{iS_{-}x}\right]$ with the actions given by $S_{\pm} = \pm(4/5)x^{5/4}$. The momentum at time $t = 0$ is then given by $p = P_{0}(x) = \partial_{x}S_{\pm} = \pm x^{1/4}$, which may be rewritten as 
\begin{equation} \label{eqn:initial_position}
    X_{0}(p) = p^{4}.
\end{equation}
Now, Hamilton's equations govern that each point on the curve evolves as
\begin{equation} \label{eqn:classical_position}
    x = X_{t}(p) = X_{0}(p) + 4p^{3}t
\end{equation}
with $X_{0}(p) = p^{4}$ given in \eqref{eqn:initial_position}. In the $x-t$ space, the trajectories \eqref{eqn:classical_position} trace lines which are shown in Fig. \ref{fig:classical_paths}. These lines are enveloped by the caustic curve $x = -27t^{4}$, which coincides with the previously obtained caustic \eqref{eq:caustic}. Another thing to note is that the classical paths also mirror the general qualitative behavior of the evolution of the $\mathrm{Ai}_{4}$ wavepacket as shown in Fig. \ref{fig:time_dynamics_abs(psi)}, particularly in that there is significant concentration near $x \approx 0$ and that it accelerates along the caustic \eqref{eq:caustic}.

\subsection{Finite Energy Initial Data}

\begin{figure}[tbhp!]
    \centering
    \includegraphics[width=0.5\textwidth]{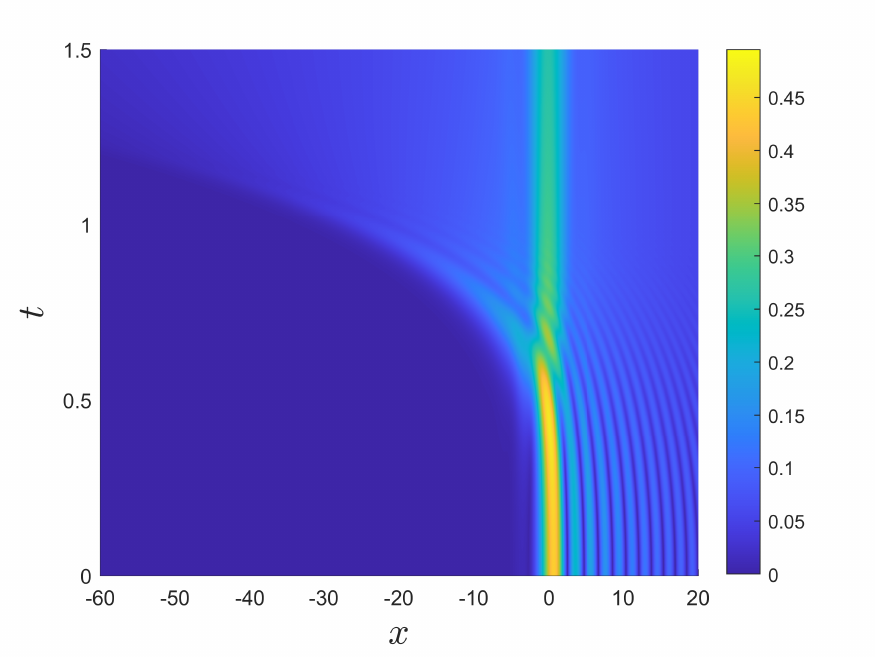}\\
    \includegraphics[width=0.5\textwidth]{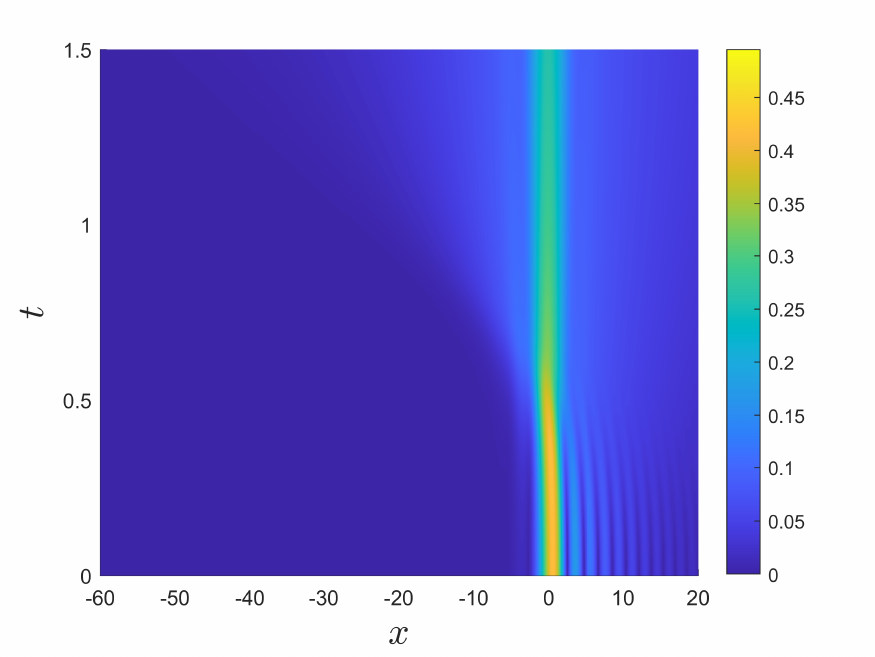}
    \caption{Propagation of the damped beam \eqref{eq:Ai4_damped_ic} with (a) $\alpha=0.02$, (b) $\alpha=0.1$.}
    \label{fig:time_dynamics_damped_hyperairy}
\end{figure}

We also consider the hyper-Airy initial profile 
\begin{equation}
\label{eq:Ai4_damped_ic}
    u(x,0)=\mathrm{Ai}_4(x)e^{-\alpha x},
    \qquad \alpha>0.
\end{equation}
The exponential factor suppresses the algebraic tail of the oscillatory
part of $\mathrm{Ai}_4(x)$ as $x\to\infty$, producing a finite-energy
beam with $u(\cdot,0)\in L^2(\mathbb{R})$. Consequently, the accelerating part of the wave packet along the caustic disperses at a much faster rate.

Taking the Fourier transform of \eqref{eq:Ai4_damped_ic} yields
\begin{equation}
    \widehat{u}(k,0)
    =
    \exp\!\left[-\frac{i}{5}(k-i\alpha)^5\right].
\end{equation}
The subsequent time evolution is obtained by multiplying it by the quartic dispersive phase factor in Fourier space and applying the
inverse Fourier transform, as in the undamped case. Figure~\ref{fig:time_dynamics_damped_hyperairy} shows the propagation for two different values of $\alpha$. When $\alpha$ is increased, the decay along the caustic is strengthened and the wave packet leaves behind only the principal lobe near the origin as time progresses.

\section{Higher Even-order Dispersions}
\begin{figure}[tbhp]
    \centering
    \includegraphics[width=0.5\textwidth]{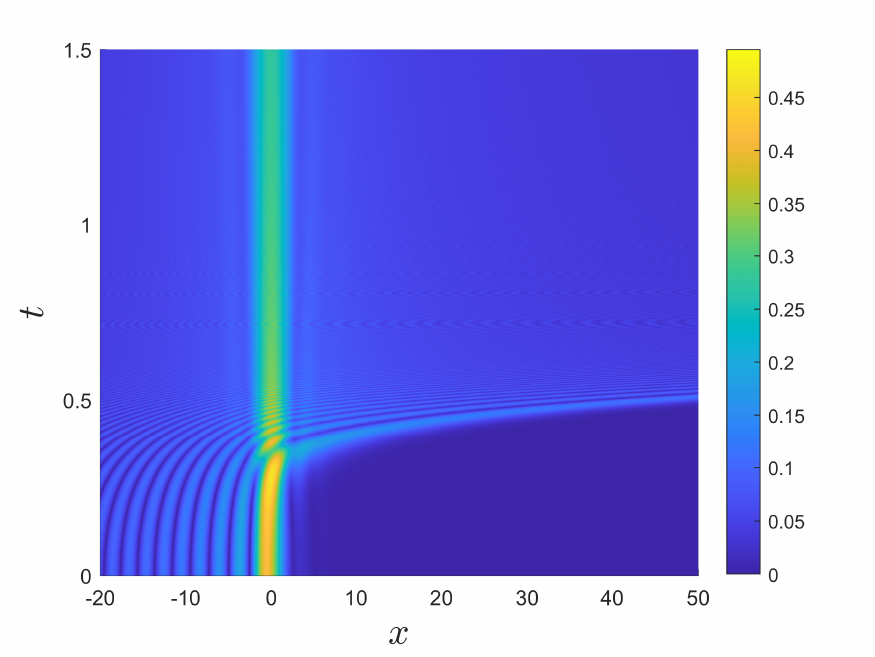}
    \caption{Propagation of $\mathrm{Ai}_6(x)$ under the free sixth-order Schr\"odinger equation. An accelerating caustic forms along $x\sim t^6$, while the amplitude decays dispersively.}
    \label{fig:time_dynamics_abs(psi)_6th}
\end{figure}

Similar behavior is observed in the free Schr\"{o}dinger equation at higher orders of dispersion. We numerically compute the propagation of a sixth-order Airy beam, $\mathrm{Ai}_{6}(x)$, which satisfy the equation
\begin{equation}
    \mathrm{Ai}_{6}^{(6)} -x \mathrm{Ai}_{6} = 0.
\end{equation}
%in the sixth-order free Schr\"{o}dinger equation. 
The dynamics of this wave packet is plotted in Fig.~\ref{fig:time_dynamics_abs(psi)_6th}. Similar to the hyper-Airy wave packet in the fourth-order Schr\"{o}dinger, the $\mathrm{Ai}_{6}(x)$ wave packet accelerates along a caustic curve, where in this case, it is $x \sim t^{6}$. Its amplitude also decreases along this caustic. To the left of this caustic, the wave packet is oscillatory and to the right, it is exponentially decaying in space. Moreover, in the vicinity of $x \approx 0$, the wave packet also leaves behind a lobe-like structure which disperses slowly in time.

For general even-order dispersion Schr\"{o}dinger equations with analogous higher order Airy beam initial conditions, we conjecture that this leads to similar propagation of the wave packets. This can be seen by doing a formal analysis using similar asymptotic methods. If we consider a Schr\"{o}dinger equation with dispersion of order $2n$:
\begin{equation}
    i \partial_{t}u = (-1)^{n}\partial_{x}^{2n}u
\end{equation}
with the initial wave packet being $u(x,0) = \mathrm{Ai}_{2n}(x)$ satisfying
\begin{equation}
    \mathrm{Ai}_{2n}^{(2n)}(x) - x \mathrm{Ai}_{2n}(x) = 0,
\end{equation}
then the wave packet would accelerate along the caustic $x \sim t^{2n}$ while dispersing over time at a rate of $O(t^{-(2n-2)/3})$. Near $x \approx 0$, the wave packet forms a principal lobe which decays at a rate of $O(t^{-1/2n})$. All other regions would decay faster than this as $t \to \infty$. Thus, the packet's amplitude decays uniformly as $O(t^{-1/2n})$ as $t \to \infty$.

\section{Conclusion and Outlook}

We have investigated whether the Schr\"odinger equation with pure quartic dispersion admits accelerating nonspreading wave packets analogous to the Berry--Balazs Airy solution of the quadratic case. By analyzing a general shape-preserving ansatz, we showed that no rigidly accelerating solutions are compatible with a local potential $V(x,t)$.
%In particular, any profile of the form $f(x-d(t))e^{i\phi(x,t)}$ necessarily satisfies $\dot d(t)=0$, so accelerated translation is excluded.

We further examined the time evolution of hyper-Airy initial data. Although quartic dispersion produces an accelerating caustic, the amplitude along this ridge decays in time according to the
dispersive scaling law $t^{-1/4}$. Thus, while structured phase concentration occurs, it does not give rise to a shape-preserving accelerating wave packet. Moreover, the principal lobe is rather stationary. The self-accelerating Airy phenomenon is therefore a special feature of quadratic dispersion and does not extend to the pure-quartic case. The same mechanism readily excludes rigidly accelerating nonspreading solutions for higher even-order dispersive operators.

Several directions merit further investigation. First, it would be natural to analyze mixed-dispersion models in which quadratic and quartic terms compete, in order to determine whether
approximate accelerating structures can arise through asymptotic balance. Second, nonlinear extensions may alter the dispersive scaling and permit accelerating localized states through dispersion--nonlinearity interplay. Finally, a systematic classification of accelerating caustics for general polynomial dispersion relations may clarify which structural features of the quadratic operator enable exact nonspreading motion.

\section*{Data availability}
No data was used for the research described in the article.

\section*{Declaration of competing interest} 
The authors declare that they have no known competing financial interests or personal relationships that could have appeared to influence the work reported in this paper.

\section*{CRediT authorship contribution statement}
The manuscript was written with contributions from all authors. All authors have given their approval to the final version of the manuscript.

\textbf{FTA}: Software, Methodology, Formal Analysis, Investigation, Writing - Original Draft; \textbf{HS}: Conceptualization, Writing - Review \& Editing.

\section*{Declaration of generative AI and AI-assisted technologies in the writing process}

During the preparation of this work, the authors used Grammarly and ChatGPT in order to improve language and readability. After using these tools/services, the authors reviewed and edited the content as needed and take full responsibility for the content of the publication.

\section*{Acknowledgements}
We would like to thank M.I.\ Yunus and the four reviewers for their suggestions, particularly on the discussion of classical paths of the hyper-Airy packet.
%HS acknowledges support by Khalifa University through the Research \& Innovation Grants under project ID KU-INT-RIG-2024-8474000789.

\bibliographystyle{elsarticle-num}
\bibliography{references}

%% Authors are advised to use a BibTeX database file for their reference list.
%% The provided style file elsarticle-num.bst formats references in the required Procedia style

%% For references without a BibTeX database:

% \begin{thebibliography}{00}

%% \bibitem must have the following form:
%%   \bibitem{key}...
%%

% \bibitem{}

% \end{thebibliography}

\end{document}